\documentclass[twocolumn]{aastex631}
\usepackage{apjfonts}

\def\aap{{\em A}\&{\em A}}

\def\apjl{{\em ApJL}}
\def\apjs{{\em ApJS}}

\newcommand{\kms}{\ifmmode {\rm km\ s}^{-1} \else km s$^{-1}$\ \fi}
\newcommand{\ergs}{\ifmmode {\rm erg\ s}^{-1} \else erg s$^{-1}$\ \fi}
\newcommand{\lb}{\ifmmode L_{\rm Bol} \else $L_{\rm Bol}$\ \fi}
\newcommand{\ledd}{\ifmmode L_{\rm Edd} \else $L_{\rm Edd}$\ \fi}
\newcommand{\lx}{\ifmmode L_{\rm 2-10keV} \else  $L_{\rm 2-10keV}$\ \fi}
\newcommand{\ha}{\hbox{H$\alpha$}}
\newcommand{\hb}{\hbox{H$\beta$}}

\newcommand{\mbh}{\ifmmode M_{\rm BH}  \else $M_{\rm BH}$\ \fi}
\newcommand{\lv}{\ifmmode \lambda L_{\lambda}(5100\Ang) \else $\lambda L_{\lambda}(5100\Ang)$\ \fi}
\newcommand{\lbol}{\ifmmode L_{\rm Bol} \else $L_{\rm Bol}$\ \fi}

\newcommand{\oi}{\hbox{[O\,{\sc i}]}}
\newcommand{\oii}{\hbox{[O\,{\sc ii}]}}
\newcommand{\nii}{\hbox{[N\,{\sc ii}]}}
\newcommand{\sii}{\hbox{[S\,{\sc ii}]}}
\newcommand{\oiii}{\hbox{[O\,{\sc iii}]}}

\newcommand{\msun}{ M_{\odot}}

\newcommand{\hii}{\hbox {H\,{\sc ii}}}

\newcommand{\oh}{\ifmmode 12+ \log({\rm O/H}) \else 12+log(O/H) \fi}
\newcommand{\mdot}{\ifmmode \dot{m} \else \dot{m} \fi }
\newcommand{\llog}{\ifmmode {\rm log} \else {\rm log} \fi }
\newcommand{\Ang}{\mathring{\mathrm{A}}}

\submitjournal{ApJL}

\begin{document}

\title{Detection of metal enrichment by SN 2011jm in NGC 4809}

\correspondingauthor{Yulong Gao, Qiusheng Gu}
\email{yulong@nju.edu.cn, qsgu@nju.edu.cn}

\author[0000-0002-5973-694X]{Yulong Gao}
\affiliation{School of Astronomy and Space Science, Nanjing University, Nanjing 210093, China}
\affiliation{Key Laboratory of Modern Astronomy and Astrophysics (Nanjing University), Ministry of Education, Nanjing 210093, China}

\author{Qiusheng Gu}
\affiliation{School of Astronomy and Space Science, Nanjing University, Nanjing 210093, China}
\affiliation{Key Laboratory of Modern Astronomy and Astrophysics (Nanjing University), Ministry of Education, Nanjing 210093, China}

\author{Ping Zhou}
\affiliation{School of Astronomy and Space Science, Nanjing University, Nanjing 210093, China}
\affiliation{Key Laboratory of Modern Astronomy and Astrophysics (Nanjing University), Ministry of Education, Nanjing 210093, China}

\author{Shi Yong}
\affiliation{School of Astronomy and Space Science, Nanjing University, Nanjing 210093, China}
\affiliation{Key Laboratory of Modern Astronomy and Astrophysics (Nanjing University), Ministry of Education, Nanjing 210093, China}

\author{Xiangdong Li}
\affiliation{School of Astronomy and Space Science, Nanjing University, Nanjing 210093, China}
\affiliation{Key Laboratory of Modern Astronomy and Astrophysics (Nanjing University), Ministry of Education, Nanjing 210093, China}


\begin{abstract}
The cosmic metals are believed to originate from stellar and supernovae (SNe) nucleosynthesis, dispersed into the interstellar medium (ISM) through stellar winds and supernova explosions. In this paper, we present the clear evidence of metal enrichment by a type Ic SN 2011jm in the galaxy NGC 4809, utilizing high spatial resolution Integral Field Units (IFU) observations obtained from the Very Large Telescope (VLT)/Multi Unit Spectroscopic Explorer (MUSE). Despite SN 2011jm being surrounded by metal-deficient ISM ($\sim 0.25 \ Z_\odot$) at a scale about 100 pc, we clearly detect enriched oxygen abundance ($\sim 0.35 \ Z_\odot$) and a noteworthy nitrogen-to-oxygen ratio at the SN site. Remarkably, the metal pollution is confined to a smaller scale ( $\la$ 13 pc). We posit that the enhanced ionized metal stems from stellar winds emitted by massive stars or previous SNe explosions. This observation may represent the first direct detection of chemical pollution by stellar feedback in star-forming galaxies beyond the Local Volume.
\end{abstract}

\keywords{galaxies: star formation - galaxies: individual: NGC4809 - galaxies:ISM - supernovae: general - stars: massive}

\section{Introduction}
\label{sec:intro}

Core-collapse supernovae (CCSNe) originate from the gravitational collapse of the iron-group element cores in massive stars ($\geq 8 \ \msun$). The historical classification of CCSNe mainly divides them into two types: hydrogen-poor (stripped-envelope) Type Ibc SNe and hydrogen-rich Type II SNe. These explosive events eject heavy elements created during core nuclear burning \citep[e.g.,][]{hoyle1960,nomoto2006,johnson2019}, leading to the chemical enrichment of the interstellar medium (ISM) \citep[e.g.,][]{timmes1995,roy2020} and potentially influencing the formation of new stars \citep[e.g.,][]{thornton1998}. CCSNe play a crucial role in the formation and distribution of chemical elements during the evolution of galaxies \citep[e.g.,][]{scannapieco2008a,kuncarayakti2013,niino2015,Kuncarayakti2018,Xiao2018,xiao2019,ganss2022,sarbadhicary2022,kourniotis2023,molero2022}.

The classification of CCSNe types reflects the different parameters of their progenitors, including initial mass, age, metallicity, and binarity. The absence of hydrogen (and helium) lines in Type Ibc SNe indicates that the progenitor star has shed its outer envelopes before the explosion. This can occur either through strong stellar winds from an evolved single massive ($M \geq 25 - 30 \ \msun$) Wolf -– Rayet (WR) star \citep[e.g.,][]{crowther2007,smith2014} or through a less massive star in a close binary system losing outer envelopes to a companion through accretion \citep[e.g.,][]{yoon2010,dessart2011}. Notably, the resolved stellar populations in the environments of Type IIb, Type Ib, and Type Ic CCSNe are very young \citep{maund2018}, i.e., log(age/yr) = 7.20, 7.05, and 6.57, respectively. Given the short lifetimes of massive stars, these CCSNe are expected to be associated with the star formation ($\hii$) regions of galaxies.

Massive stars are particularly relevant to galaxy evolution, especially at low metallicity regimes, as they significantly contribute to the chemical enrichment of the ISM through the loss of metals (e.g., C, N, O) via strong stellar winds \citep[e.g.,][]{dray2003,smith2014,luo2021a}. Furthermore, the final SNe of these stars can lead to significant enrichment in certain elements, particularly oxygen.

However, an intriguing contradiction exists between the metal enrichment of ISM by stellar activities and observations of metallicity and star formation rates (SFRs) in galaxies. Analyses using single spectra from the Sloan Digital Sky Survey (SDSS) by \cite{Mannucci2010} and \cite{Andrews2013} revealed an anticorrelation between gas-phase metallicity and SFR at a fixed stellar mass in galaxies, suggesting that galaxies undergoing more intense star formation activities appear to be metal deficient. Similarly, this trend has also been detected in stellar mass surface density, SFR surface density, and local metallicity based on Integral Field Units (IFU) observations of star-forming galaxies \citep{Cresci2019,hwang2019,curti2020,yao2021a}. They found that the oxygen abundance (12 + log(O/H)) in intense star formation regions is significantly lower than the tight empirical relation between metallicity and local stellar mass surface density (spatially resolved MZR) \citep[e.g.,][]{Barrera-Ballesteros2016a,Sanchez2017a,gao2018b}.  Recently, \cite{gao2023} studied star-forming knots in the nearby merging dwarf galaxies NGC 4809/4810 using high spatial resolution observations from the Very Large Telescope (VLT)/Multi Unit Spectroscopic Explorer (MUSE) and found that knots with higher SFRs exhibit lower metallicity.

The observed metallicity in ISM reflects the long-term interplay between enrichment from stars, metal loss caused by winds, and the dilution of gas influx \citep{wang2022p}. Calculated SFRs trace the average newly formed stars at a much shorter timescale (10 -- 100 Myr) \citep{Kennicutt2012}. The anticorrelation between metallicity and SFR can be explained by gas accretion, where the accretion of primal gas from ISM or intergalactic medium (IGM) dilutes pre-existing metal-rich gas and triggers star formation.  However, direct observational evidence of significant metal enrichment due to intense stellar activities remains scarce outside the Milky Way.

In this letter, we aim to detect metal enrichment at the site of SN 2011jm through high spatial resolution IFU observations from VLT/MUSE. SN2011jm, classified as a Type Ic supernova with an initial star mass of 20.7 $\msun$ \citep{howerton2011a,Kuncarayakti2018}, is located in the metal-poor regions of the merging dwarf galaxies NGC 4809/4810 \citep{gao2023}. The redshift ($z$) is 0.00326\footnote{obtained from NED, \url{http://ned.ipac.caltech.edu}}, corresponding to a luminosity distance of 14.0 Mpc and a scale of 67 pc per arcsecond. We will compare the oxygen and nitrogen abundance at the SNe site with their surrounding ISM, offering a more detailed understanding of metal enrichment from massive progenitor stars and SNe.

This paper is organized as follows. In Section 2, we present the observations and data reduction and then derive the ionized gas properties. The main results and discussion are presented in Sections 3 and 4, respectively. We summarize our findings in Section 5. Throughout the paper, we assume a flat $\Lambda$CDM cosmology model with $\Omega_\Lambda=0.7$, $\Omega_{\rm m}=0.3$, and $H_0=70$ km s$^{-1}$ Mpc$^{-1}$. We adopt the solar metallicity ($Z_{\odot}$) as $\oh = 8.69$ \citep{asplund2009}.

\section{Observations and data reduction}
\label{sec:data}

\begin{figure*}[t]
   \centering
   \includegraphics[width=0.36\textwidth]{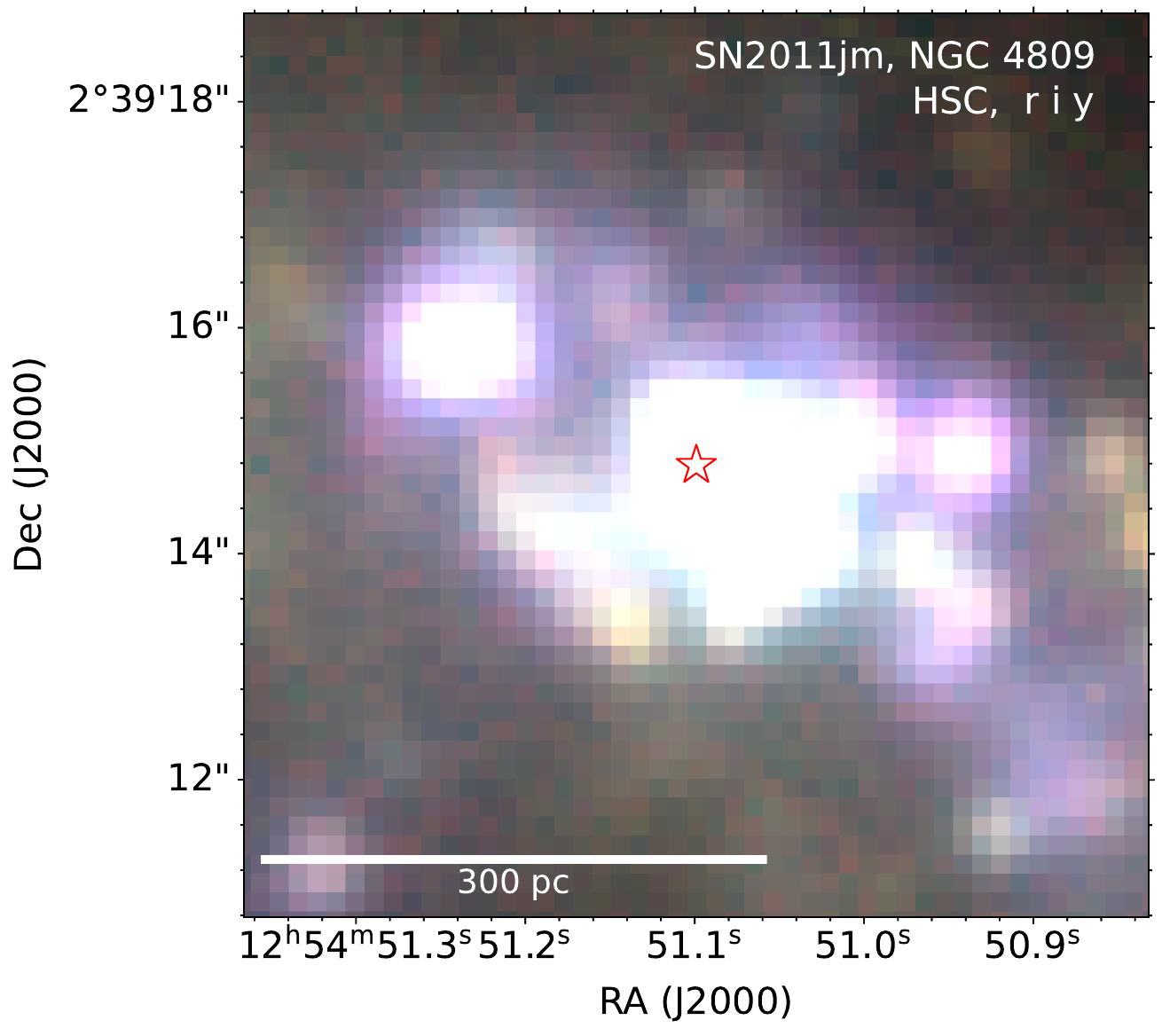}
   \includegraphics[width=0.45\textwidth]{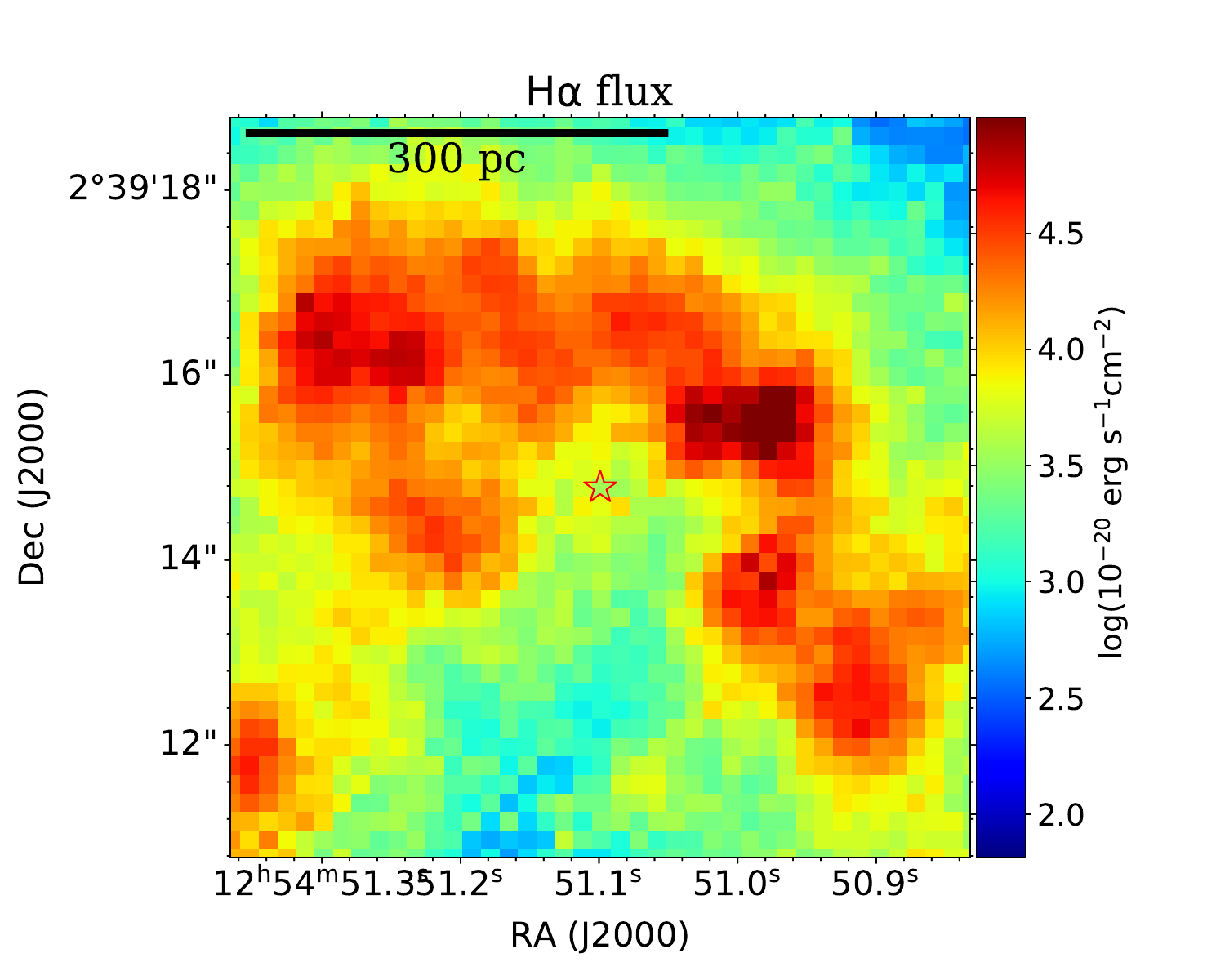}
   \includegraphics[width=0.45\textwidth]{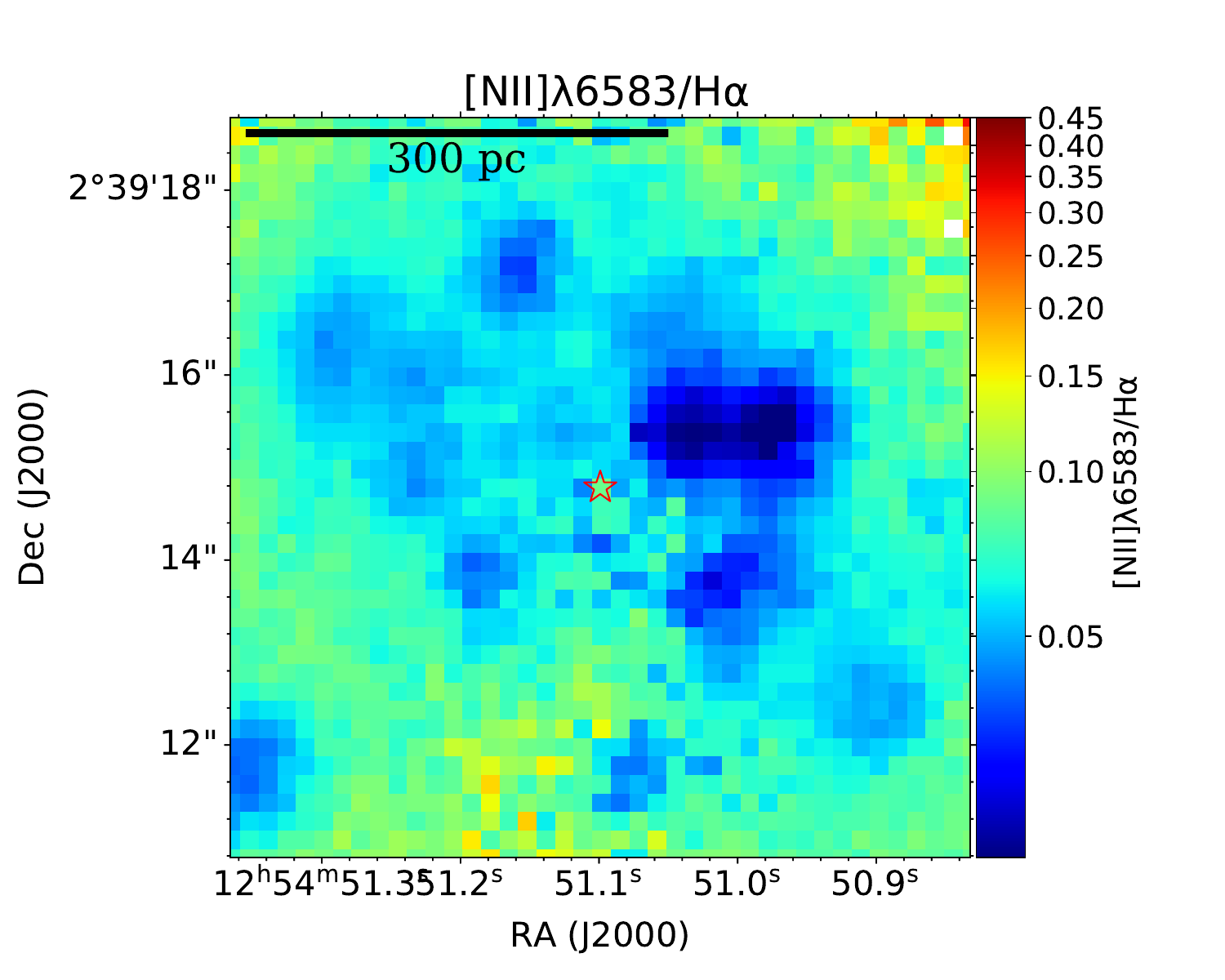}
   \includegraphics[width=0.45\textwidth]{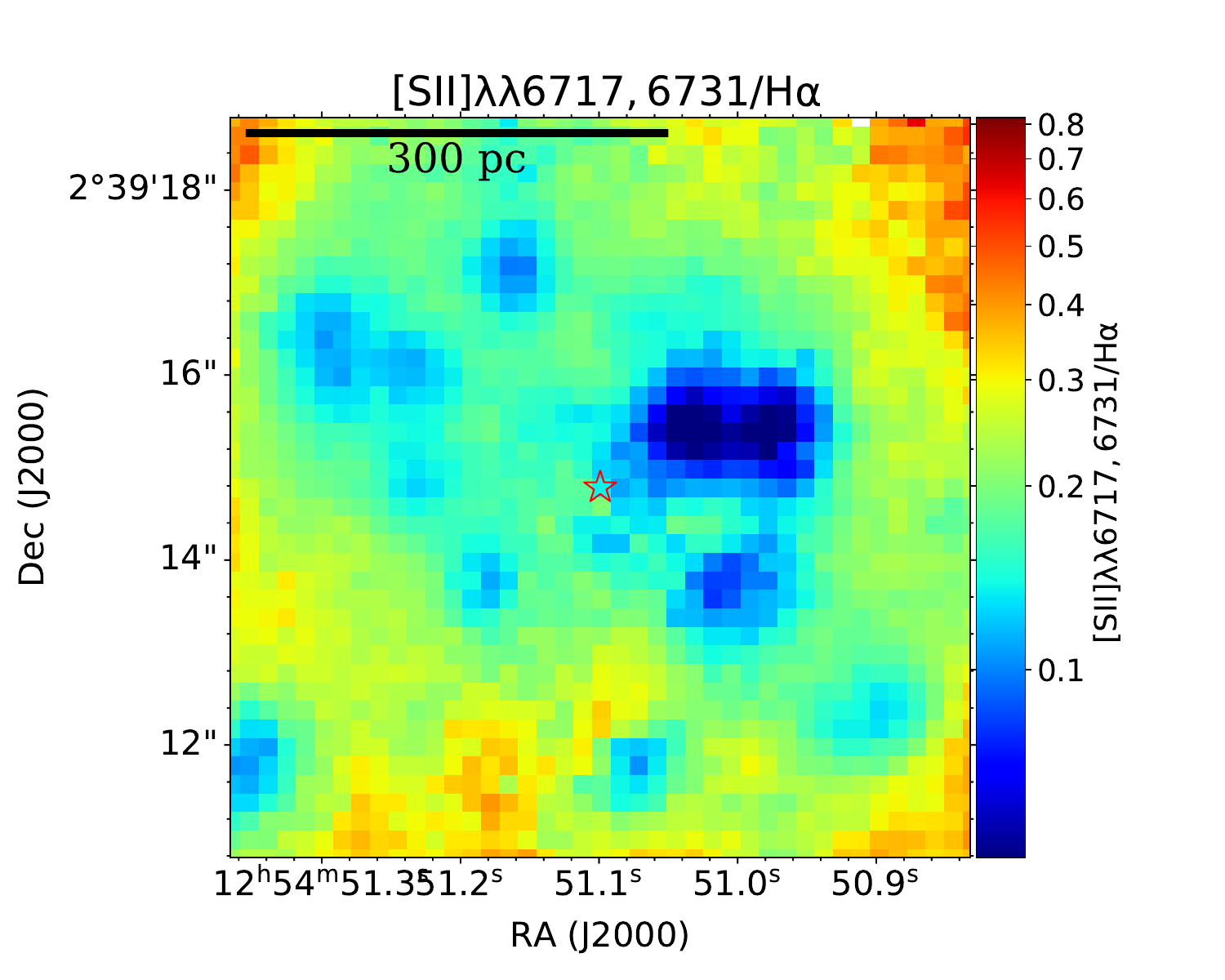}
   \caption{$Upper$: Pseudo-color image of $8\arcsec \times 8\arcsec$ region around the SN 2011jm (red star) in dwarf galaxy NGC 4809 combined with the $r$-, $i-$, and $y$ -band images from the HSC-SSP (left panel). Integrated-intensity map of the attenuation-corrected $\ha$ emission (right panel). $Bottom$: The intensity ratio $\nii\lambda6583/\ha$ (left) and $\sii\lambda\lambda6717,6731/\ha$ (right) maps. The red stars represent the SN 2011jm site. } 
   \label{fig:Ha}
 \end{figure*}

NGC 4809/4810 underwent observation with VLT/MUSE in May 2015 (ID: 095.D-0172; PI: Kuncarayakti), featuring an integration exposure time of approximately 0.8 hours within the two-component field-of-view regions. The fully reduced data cube was sourced from the ESO archive website\footnote{\url{http://archive.eso.org/scienceportal/home}}, with a pixel scale of 0.2\arcsec. The observations achieved a seeing value of about 0.5 $\arcsec$. The full width at half maximum (FWHM) of the final image is approximately 0.8$\arcsec$. Covering a rest-frame spectral range of $\rm 4750 - 9160 \Ang$, with a channel width of 1.25$\rm \Ang$, the data cube provided a comprehensive dataset. The upper-left panel of Fig. \ref{fig:Ha} showcases a Pseudo-color image of the $8\arcsec \times 8\arcsec$ region around SN 2011jm in NGC 4809, combining $r$-, $i-$, and $y$ -band images from the Hyper Suprime-Cam Subaru Strategic Program (HSC-SSP).

We employed the STARLIGHT package \citep{CidFernandes2005}, assuming the \cite{Chabrier2003} initial mass function (IMF), and conducted a synthesis of the stellar continuum using a combination of 45 single stellar populations (SSPs) from the \cite{Bruzual2003} model. The subtraction of the stellar continuum synthesis was followed by the application of multiple Gaussian fits to capture strong emission lines, including $\ha$, $\hb$, $\oiii\lambda\lambda4959,5007$, $\oi\lambda6300$, $\nii\lambda\lambda6548,6583$, and $\sii\lambda\lambda6717,6731$. In our pursuit of a reliable measurement of metallicity, we restricted our analysis to spaxels with a signal-to-noise ratio (S/N) exceeding 5 for these emission lines. In Fig. \ref{fig:Ha}, we also show the maps of integrated-intensity of attenuation-corrected $\ha$ emission, and the intensity ratios of $\nii\lambda6583/\ha$ and $\sii\lambda\lambda6717,6731/\ha$. The ratios of $\nii\lambda6583/\ha$ and $\sii\lambda\lambda6717,6731/\ha$ can trace the gas-phase metallicity and shocked gas, respectively. 

Typically, SNe explosion in dense environments manifest their presence through narrow emission lines evident in optical bands. These collisionally excited lines, such as the $\nii\lambda6583/\ha$ and $\sii\lambda6731/\ha$ line ratios, are crucial for classifying an object as a SN remnant (SNR). \cite{makarenko2023} proposed a criterion, requiring $\nii\lambda6583/\ha > 0.5$ and $\sii\lambda6731/\ha > 0.4$, to identify SNRs. However, at the pixel of SN 2011jm site, we compute the ratios of $\nii\lambda6583/\ha = 0.12 \pm 0.02$ and $\sii\lambda6731/\ha = 0.06 \pm 0.01$, significantly below the specified criterion. Additionally, the $\oiii\lambda5007/\hb$ ratio is about 2.13, indicating that the SN 2011jm site is situated in a star-forming region on the $\nii$-based and $\sii$-based BPT diagrams \citep{Baldwin1981,Kauffmann2003,Kewley2001,kewley2006}. These findings suggest that the observed emission at the SN site is primarily a result of photoionization in $\hii$ regions rather than shocked gas.

\section{Results}
\label{sec:results}

\subsection{Dertermination of oxygen and nitrogen abundence}
\label{subsec:metal_cali}

\begin{figure*}[t]
   \centering
   \includegraphics[width=0.45\textwidth]{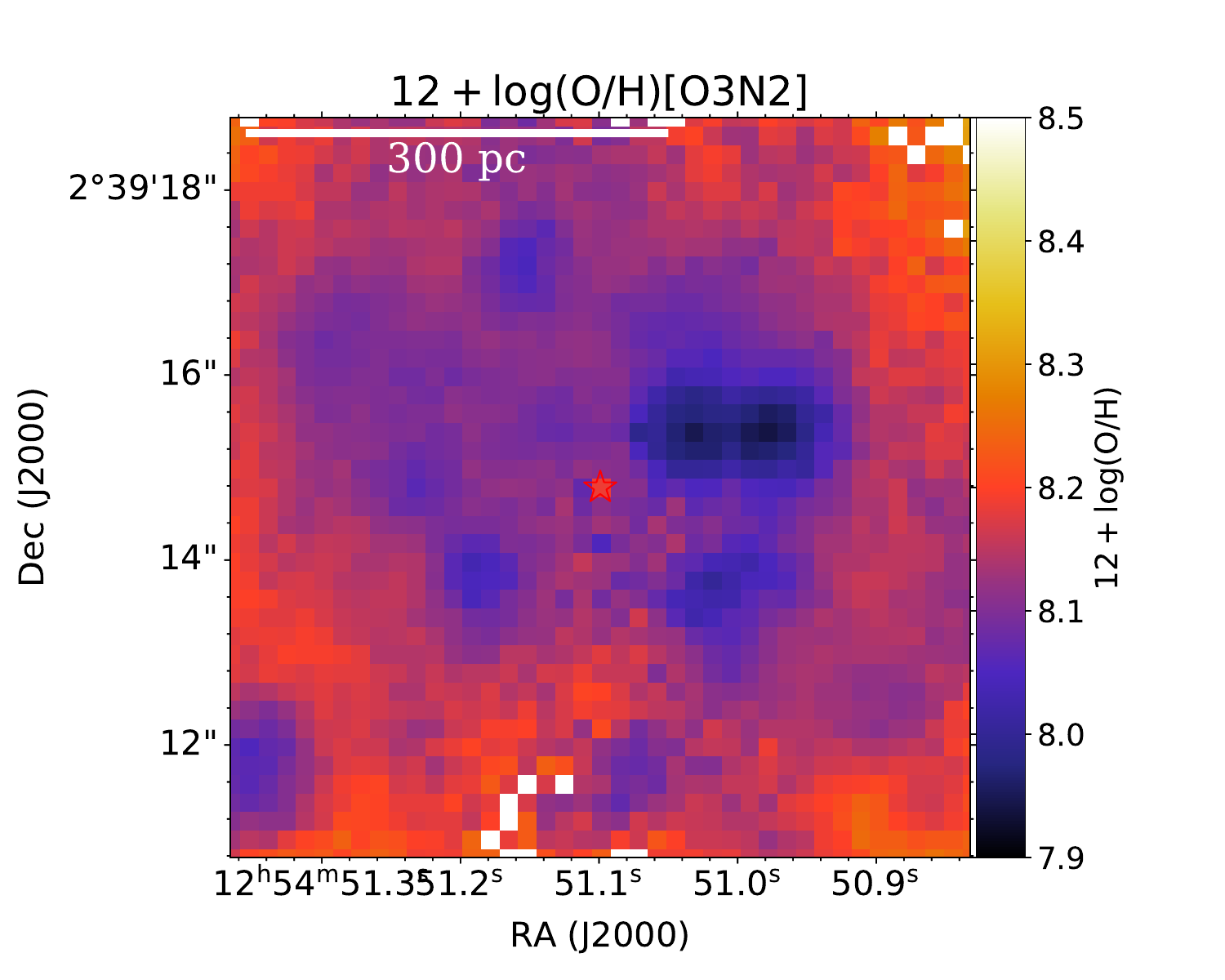}
   \includegraphics[width=0.45\textwidth]{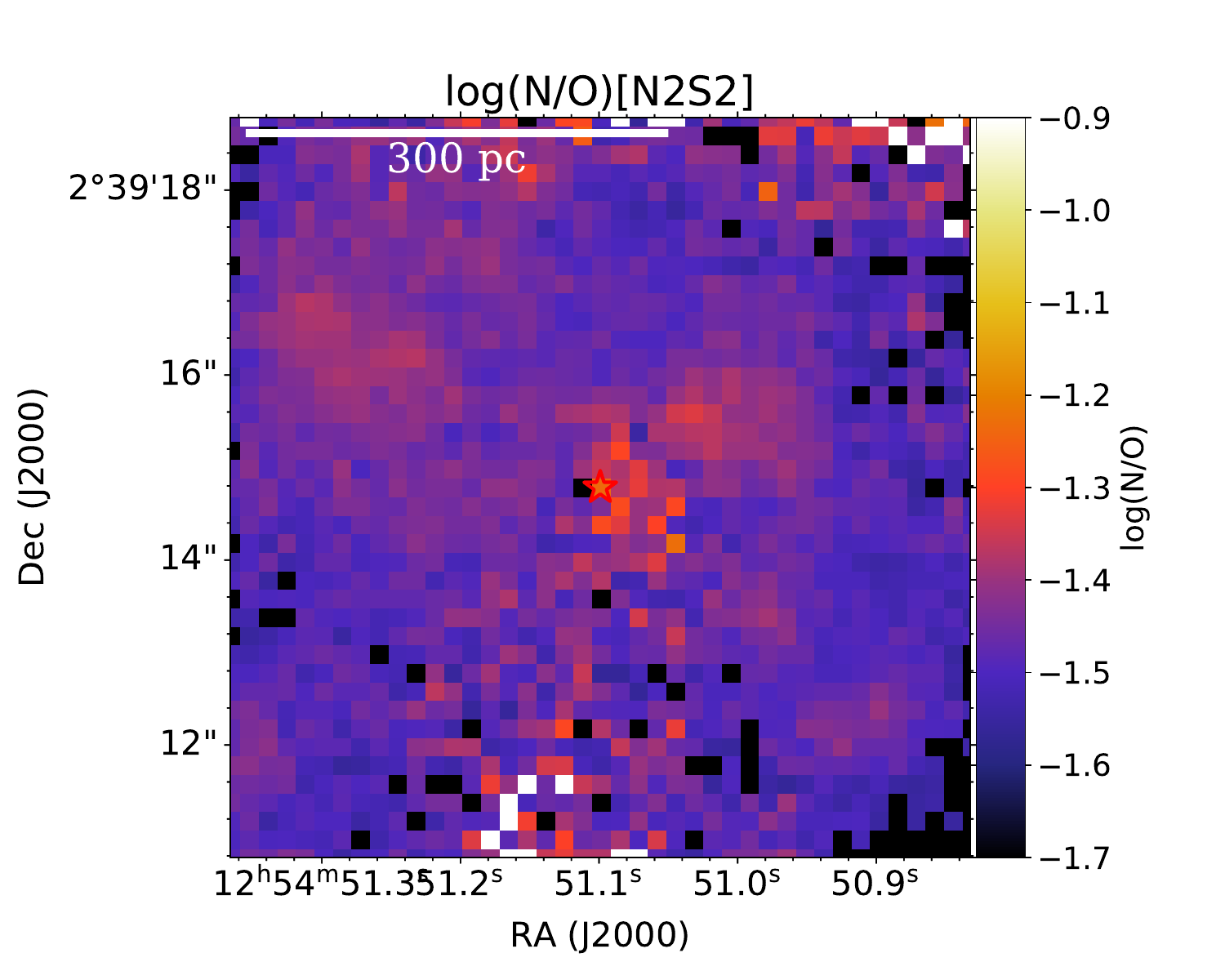}
   \caption{Gas-phase oxygen abundance (left) and nitrogen to oxygen ratio (N/O, right) maps around the SN 2011jm.} 
   \label{fig:zoh_no_map}
 \end{figure*}

Understanding the gas-phase metallicity in galaxies is pivotal for unraveling the intricacies of their evolutionary processes. Oxygen, as a primary element, holds a special significance due to its independence from the initial chemical composition of stars. Helium burning in massive stars produces oxygen, ejected into the ISM via CCSNe on short timescales \citep[e.g.,][]{dray2003}.
In contrast, nitrogen exhibits both primary and secondary origins. Primarily produced through the CNO cycle in low- and intermediate-mass stars ($< 8 \ \msun$), nitrogen is ejected into the ISM during the post-main-sequence asymptotic giant branch (AGB) phase. In low-metallicity stars, nitrogen acts as a primary element, maintaining a fixed ratio of N/O as the required carbon and oxygen were created in the star via helium burning. Conversely, in high-metallicity stars, nitrogen behaves as a secondary element, with N/O increasing with O/H as most of the carbon and oxygen used to form nitrogen existed in the pre-existing ISM \citep{luo2021a}.

Several methodologies exist for measuring gas-phase metallicity, particularly oxygen abundance, in the ISM \cite[e.g.,][]{kewley2008,kewley2019}. Photoionization models of $\hii$ regions are employed to reproduce specific emission line ratios, such as N2O2 ($\nii\lambda6583/\oii\lambda3727$) \citep{kewley2002}, R23 ($(\oii\lambda3727+\oiii\lambda\lambda4959,5007)/\hb$) \citep{kobulnicky2004}, and N2S2 ($\nii\lambda6583/\sii\lambda\lambda6717,6731$) \citep{dopita2016}. The most reliable approach involves directly measuring the electron temperature ($T_{\rm e}$) from faint auroral-to-nebular emission line ratios, such as $\oiii\lambda4363$/$\oiii\lambda5007$. This method, however, is applicable mainly in $\hii$ regions with high temperatures and minimal cooling in metal-poor galaxies, a condition not uniformly met. As the MUSE spectral wavelength does not cover the $\oii\lambda3727$ and $\oiii\lambda4363$ lines, we employed the strong-line calibrator, O3N2, to ascertain the oxygen abundance following the calculation in \cite{xiao2019} and \cite{ganss2022}. 

The O3N2 index \citep{Alloin1979} is defined as
\begin{equation}
\rm O3N2 \equiv \log\left(\frac{\oiii\lambda5007}{\hb} \times \frac{\ha}{\nii\lambda6583}\right).
\end{equation}
\cite{Marino2013} improved the O3N2 calibration based on CALIFA and literature data using $T_{\rm e}$, providing the following relation:
\begin{equation}
\rm 12 + \log(O/H) = 8.505 - 0.221 \times O3N2
,\end{equation}
when O3N2 ranges from $-1.1$ to 1.7. The median O3N2-based metallicity and its 1$\sigma$ dispersion of all spaxels are about $8.10 \pm 0.03$. 

We also obtain the nitrogen-to-oxygen ratio using the N2S2 index \citep{dopita2016}, which is defined as
\begin{equation}
\rm N2S2 \equiv \log\left(\frac{\nii\lambda6583}{\sii\lambda\lambda6717,6731}\right).
\end{equation}
\cite{florido2022} provided the new empirical calibration of nitrogen to oxygen ratios, showing as follows:
\begin{equation}
\rm \log(N/O) =  0.84 \times N2S2 - 1.071.
\end{equation}
The average N2S2-based log(N/O) and its 1$\sigma$ dispersion are about $-1.45 \pm 0.04$. 

We conduct a comprehensive analysis by repeating the emission line fitting and calculation process 10,000 times to accurately estimate the uncertainties in the determination of metallicity and the N/O ratio. In each iteration, we add the noise from the continuum spectra into the spectral intensity of each emission line. This noise is generated using random numbers from a Gaussian distribution with a mean of zero and a standard deviation. The deviation is determined as the Root-Mean-Square (RMS) value within 100$\Ang$ around each emission line. The median value among the 10,000 measurements is adopted as the representative measurement, and the 1$\sigma$ value of the distribution is combined with the calculated uncertainty derived from the observational uncertainties of the emission lines. This approach allows us to account for variations and potential systematic errors, providing a robust assessment of the metallicity and N/O ratio determinations.

The spatial distributions of gas-phase oxygen abundance and N/O ratio are illustrated in Figure \ref{fig:zoh_no_map}. We classify values below (above) 1$\sigma$ as depleted (enriched) for 12 + log(O/H) and log(N/O). The oxygen abundance distribution exhibits heterogeneity in the vicinity of SN2011jm, encompassing both oxygen-poor and oxygen-rich ionized gas. Remarkably, the SN 2011jm site displays elevated oxygen abundance and N/O ratio compared to its surrounding ISM.
If alternative other strong emission line ratio calibrator, such as N2 \citep{pettini2004} and RS32 \citep{curti2020}, were employed to estimate metallicity, the results for enriched oxygen and nitrogen abundance at the SNe site, as well as the observed decreasing trend in Section \ref{subsec:metal_profile}, align closely with those derived in this Section.

\begin{figure*}[t]
   \centering
   \includegraphics[width=0.45\textwidth]{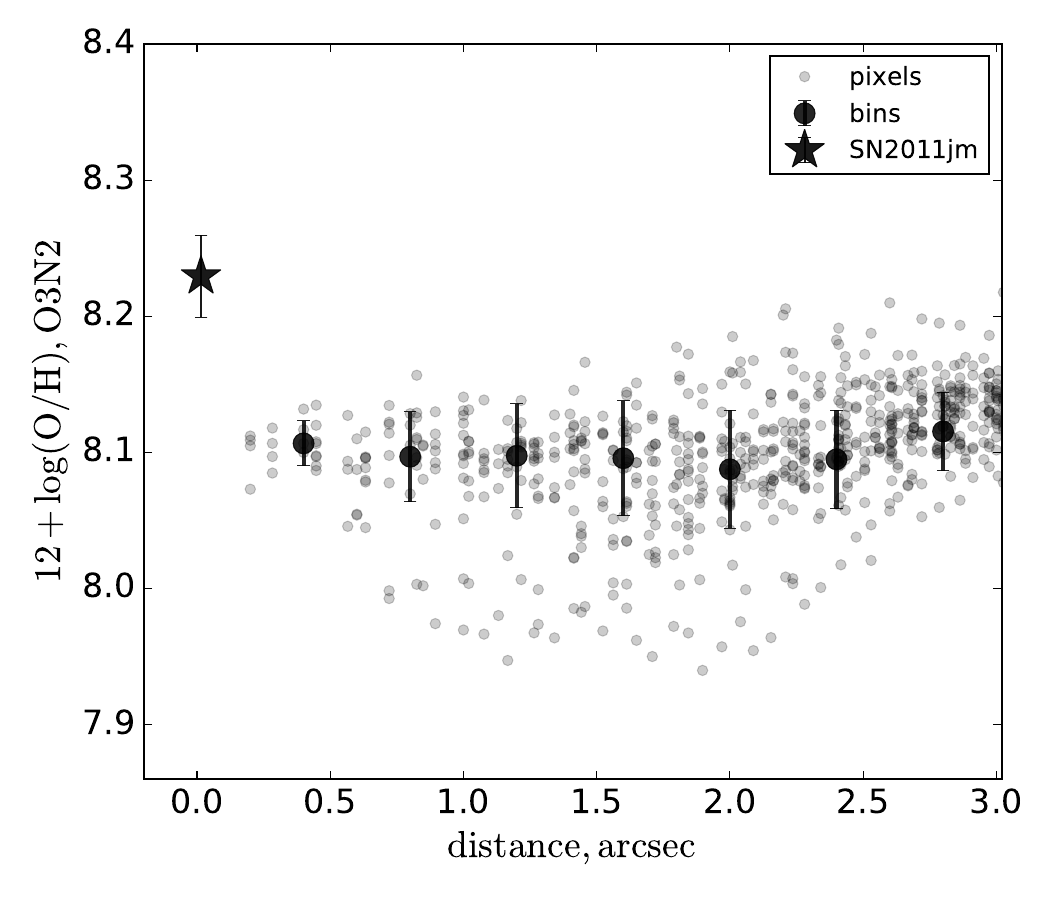}
   \includegraphics[width=0.45\textwidth]{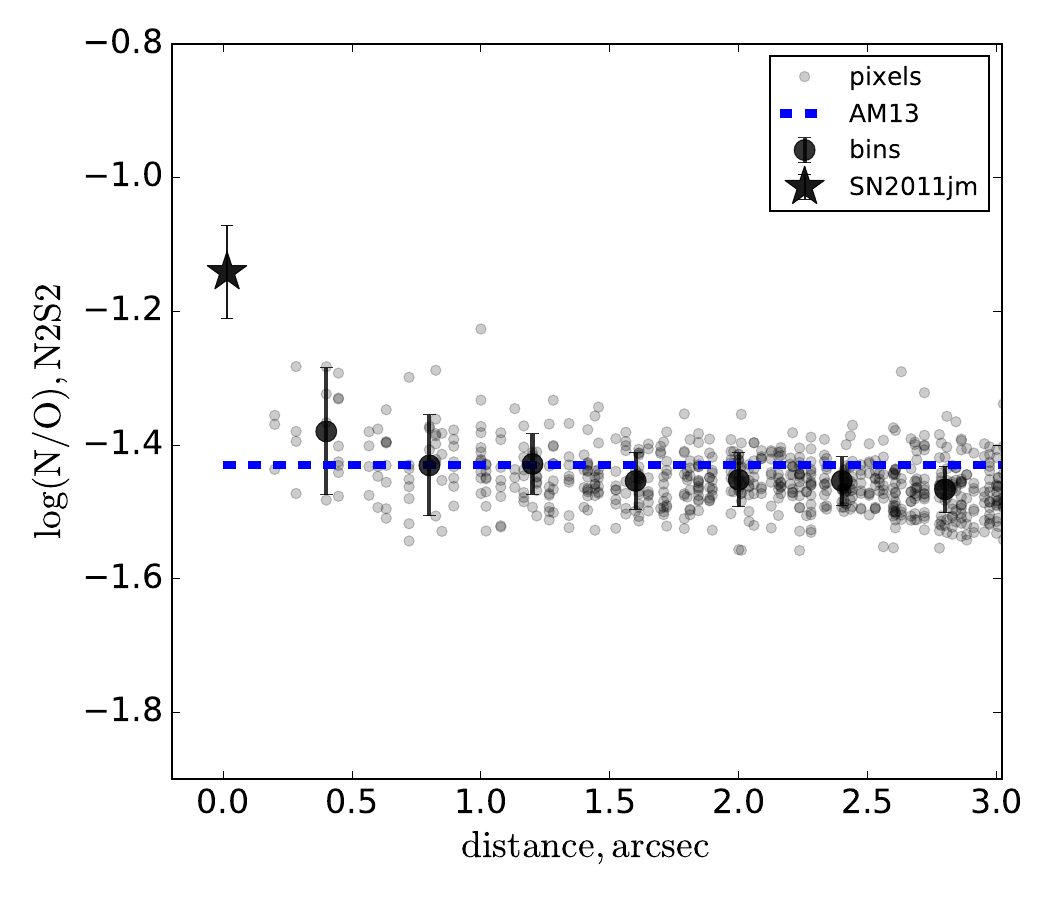}
   \caption{The left panel shows the oxygen abundance (O/H) versus the distance of pixels from the SN 2011jm site, with gray points representing values for all pixels. Distance values are grouped into 7 bins, with average values denoted by black points and 1$\sigma$ distributions marked by error bars. The black solid line represents the best-fitted relation between average O/H values and distances. In the right panel, a similar representation is shown for N/O ratios. The blue dashed line indicates the constant value of log(N/O) $\sim -1.43$ when 12 + log(O/H) < 8.5 \citep{Andrews2013}.}
   \label{fig:zoh_no_profile}
 \end{figure*}

\subsection{Metal profile}
\label{subsec:metal_profile}

To illustrate the enrichment of oxygen and nitrogen abundance, we present the O/H and N/O profiles plotted against the distance from pixels to the SN 2011jm site in Fig. \ref{fig:zoh_no_profile}. The SN 2011jm site is marked with stars, while gray points represent all pixels around the supernova at distances less than 5$\arcsec$. Black points with error bars denote the average values and their 1$\sigma$ distribution of O/H and N/O in 7 distance bins with a stepsize of $0.4\arcsec$. We observe a decreasing trend in oxygen abundance and the N/O ratio with increasing distance, particularly within 1.5$\arcsec$ ($\sim$ 100 pc). The SN 2011jm site exhibits the highest values of (12 + log(O/H) $= 8.23 \pm 0.03$) and N/O ratio (log(N/O) $= -1.14 \pm 0.07$), surpassing the 1$\sigma$ distribution. These trends suggest that the winds from progenitor massive stars and the supernova explosion contribute to the enrichment of oxygen and nitrogen abundance in the surrounding ISM.

The blue dashed line in the right panel represents the constant value of log(N/O) $\sim -1.43$ when 12 + log(O/H) < 8.5 \citep{Andrews2013}. Previous studies \citep[e.g.,][]{luo2021a, xu2021a, yao2021a} suggested that higher N/O values in low O/H regions support the scenario that accretion of metal-poor gas can sustain star formation and create more secondary nitrogen elements. However, we observe that the SN 2011jm site exhibits the highest N/O value even at the highest O/H value when the distance is less than 1.0$\arcsec$. This result is inconsistent with the previous scenario, leading us to propose that nitrogen is primarily produced by strong stellar winds from massive stars and SN explosions.

\section{Discussion}
\label{sec:dis}

\begin{figure*}[t]
   \centering
   \includegraphics[width=0.95\textwidth]{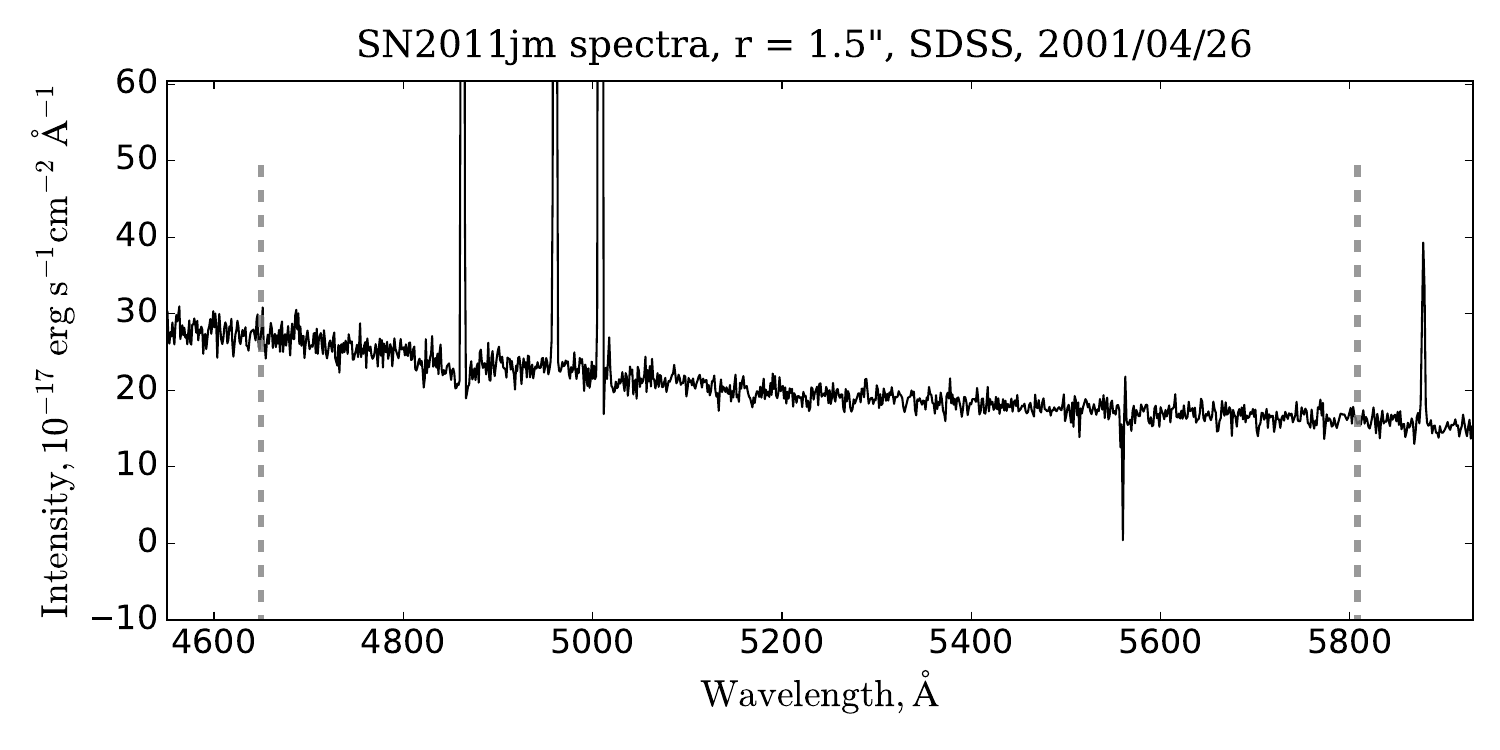}
   \caption{Spectra from a region with a size of $r = 1.5\arcsec$ around the SN 2011jm site, as observed by SDSS on 2001/04/26. The gray dashed lines indicate the absence of WR features around 4650$\Ang$ (WN stars) and 5808$\Ang$ (WC stars).}
   \label{fig:sdss_spec}
 \end{figure*}

\subsection{Progenitors of SN 2011jm}
\label{subsec:prog}

In the stellar evolution model of a single star, a Type Ic supernova can be produced by very massive stars (> 40 $\msun$) of O-type \citep{crowther2007}. These massive O stars lose substantial mass through stellar winds, revealing first the products of hydrogen burning at their surface and later the products of helium burning. These stages are spectroscopically identified with the nitrogen sequence and carbon sequence WR stars, respectively. Key WR features in their spectra include bumps around 4650$\Ang$ (i.e., the blue bump, WN stars) and 5808$\Ang$ (i.e., the red bump, WC stars). In Fig. \ref{fig:sdss_spec}, we display the SN 2011jm spectra at a region size of $r = 1.5\arcsec$, observed by SDSS on 2001/04/26, approximately ten years before the supernova explosion. However, we did not detect the WN or WC features. Furthermore, the initial mass of SN 2011jm is about 21 $\msun$, considerably smaller than the initial mass of a Type Ic supernova evolved from a single massive star. 

Another scenario is that most Type Ibc SNe evolve from moderately massive interacting binary stars \citep{smartt2009, sen2022}. In a close binary system, stars can transfer their hydrogen-rich envelope and angular momentum to a companion star during Roche lobe overflow (RLOF) before core collapse, leading to supernovae. \cite{smartt2009} presents a possible evolutionary scenario involving an interacting binary system with two stars (8 -- 30 $\msun$) that can create a Type Ic supernova, consistent with the initial mass of SN 2011jm. Consequently, we argue that the reliable progenitor of SN 2011jm is massive binary star system.

\subsection{Metal enrichment}
\label{subsec:metal_enrich}

In the local universe, many star-forming regions in galaxies exhibit lower oxygen abundance and enhanced N/O ratios. The popular explanation is that the accretion of primal gas can dilute the oxygen abundance, sustain star formation, and create more nitrogen elements. However, in the very young starburst regions of metal-poor dwarf galaxies, associated with short-lived massive stars gathered in large clusters, chemical enrichment via strong stellar winds and supernovae might be significant.

As mentioned in Section. \ref{sec:results}, we detect the enriched oxygen and nitrogen at the SN 2011jm site. Limited to the spatial resolution of the observation, the metallicity enrichment is mainly concentrated on the center pixel. Here, we assume the injection width ($w_{\rm inj}$), which represents the size of the region into which supernovae and massive stars directly deposit metals, as the pixel size, with a diameter of $0.2\arcsec \sim 13$ pc. This scale is nearly similar to the findings in the nearby galaxies \citep{lopez-sanchez2011,sarbadhicary2023}. \cite{lopez-sanchez2011} performed the IFU observations of the intense star-forming region [HL90] 111 within the starburst galaxy IC 10, and found the similar chemical pollution at a scale of 7.8 pc. \cite{sarbadhicary2023} used the Jansky VLA and Atacama Large Millimeter/submillimeter Array (ALMA) telescopes to obtain high-resolution maps of the atomic and molecular gas around WR stars and supernovae, discovering a 20 pc diameter molecular gas cavity around a WR star. Additionally, recent analyses by \cite{li2023c} and Li et al. (2023, in communication) examining 2D oxygen abundance distributions using two-point correlation functions indicate the average injection widths of about 34 pc in dwarf galaxies. If we assume the time scale of metal-rich stellar winds from (super-)AGB stars is 0.1 Myr \citep{karakas2014}, the average velocity of winds is about 60 $\kms$. \cite{decin2020} observed a sample of binary AGB stars with the ALMA telescope and found the velocities of molecular-phase winds ranging from 14 to 65 $\kms$. Our result is consistent with the findings of \cite{decin2020}. 

Here, we present a preliminary estimation of the metal mass by utilizing the $\sii$ emission line ratio and determining the electron density, approximately 50 $\rm cm^{-3}$, through the $Pyneb$ package \citep{luridiana2015}. We make the assumption that the ionized hydrogen density is equal to the electron density. Within the region with a radius of 6.5 pc, we find the ionized hydrogen mass to be around 1.4 $\times 10^3 \ \msun$, and the oxygen mass to be approximately 3.6 $\msun$ (12+log(O/H) $\sim$ 8.2). Assuming the pre-existing ISM around the supernova possesses a similar oxygen abundance to the outer region (12+log(O/H) $\sim$ 8.1), the pre-existing ionized oxygen mass is estimated to be about 2.8 $\msun$. Consequently, the newly created ionized oxygen mass is approximately 1.0 $\msun$. \cite{sukhbold2016} calculated the produced oxygen mass in the evolution of massive stars, estimating around 0.03 $\msun$ carried by stellar winds and 2.5 $\msun$ ejected by supernovae, respectively, at an initial mass of 21 $\msun$. Recently, \cite{farmer2023} performed nucleosynthesis on binary-stripped stars, finding that binaries can produce more elemental mass than single stars due to efficient mass loss and an increased chance of ejecting their envelopes during a supernova. These findings suggest two possible scenarios: one is that the enriched ionized oxygen was contributed by stellar winds from about 33 massive stars. Another is that the ejected oxygen mass from one Ic SN is enough to pollute the ISM.

However, the time between supernovae and observation is approximately five years, indicating that the pollution scale is much smaller than 1.5 pc. The FWHM value of the image is about 0.8$\arcsec$, corresponding to a diameter of $\sim$52 pc. Within this scale, it is hypothesized that stellar winds from over 300 massive stars or ejections from four Ic supernovae would be necessary. This contradicts the detection of only one supernova at this specific site. While the observations of enriched metals may be influenced by the point spread function, the affected area is notably concentrated at the central pixel.

Although we detect enriched gas-phase metallicity in the environment of the supernova site. At a larger scale ($\sim 1\arcsec$, 67 pc), the total metal is still deficient in these intense star-forming regions. This result might indicate that the dilution of primal gas accretion, instead of stellar winds or supernovae, dominates the gas-phase metallicity distribution at a $\sim$ 100 pc scale in star-forming galaxies. Consequently, our scenario can elucidate the contradiction between metal enrichment by stellar activities and the observed anticorrelation between spatial metallicity and SFRs in galaxies.

\section{Summay}
\label{sec:summay}

In this study, we explored the ionized gas-phase metallicity surrounding the type Ic SN 2011jm within the merging dwarf galaxy NGC 4809, utilizing VLT/MUSE IFU data. Our observations revealed that SN 2011jm is encircled by a metal-poor ISM with an abundance of 12 + log(O/H) around 8.1, extending over a broad scale of 100 pc. Intriguingly, we identified enriched oxygen abundance (12 + log(O/H) = 8.23 $\pm$ 0.03) and an elevated nitrogen-to-oxygen ratio (log(N/O) = $-1.14 \ \pm$ 0.07) at the supernova site. The absence of WR features in the progenitor's spectra suggests a potential association with a massive binary star system.

We propose that the enriched ionized metals may originate from either stellar winds emitted by massive stars or oxygen ejected from SN 2011jm. The diameter of the metal pollution extent is estimated to be less than 13 pc. Our findings suggest that, in star-forming galaxies, the distribution of gas-phase metallicity at a scale of approximately 100 pc is primarily influenced by the dilution of primal gas accretion rather than the impact of stellar winds or supernovae.

This study presents a noteworthy contribution as it potentially represents the first direct detection of chemical pollution resulting from stellar feedback in star-forming galaxies beyond the Local Volume. The implications of our work may provide valuable insights into reconciling the apparent contradiction between metal enrichment through stellar activities and the observed anticorrelation between spatial metallicity and SFRs in galaxies. 
However, the spatial resolution of VLT/MUSE is not enough to resolve the structure of ejected metal from SNe in galaxies beyond the Local Volume, the higher resolution ($\la 0.1\arcsec$) observations, such as JWST, are needed. 

\begin{acknowledgments}
Y.L.G acknowledge the grant from the National Natural Science Foundation of China (No. 12103023).
This work is supported by the National Natural Science Foundation of China (No. 12192222, 12192220, 12121003, and 12273010). 
We acknowledge the science research grants from the China Manned Space Project with NO. CMS-CSST. 
\end{acknowledgments}

%

\vspace{5mm}
\facilities{VLT/MUSE}


\software{Astropy \citep{astropy:2013, astropy:2018, astropy:2022}, Python, STARLIGHT \citep{CidFernandes2005}, Pyneb \citep{luridiana2015}, Matplotlib \citep{Hunter2007}}

\bibliography{sn2011jm}{}
\bibliographystyle{aasjournal}
\end{document}